\documentclass[11pt]{article}

\usepackage{amsfonts}
\usepackage{amsmath}
\usepackage{amssymb}
\usepackage{amsthm}
\usepackage[mathscr]{eucal}
\usepackage{bbold}
\usepackage{braket}
\usepackage{color}
\usepackage{comment}
\usepackage{dsfont}
\usepackage{enumitem}
\usepackage{framed}
\usepackage{jheppub}
\usepackage{mathtools}
\usepackage{physics}
\usepackage[normalem]{ulem}
\usepackage{tensor}
\usepackage{thmtools}
\usepackage{thm-restate}
\usepackage{ulem}
\usepackage{tikz}
\usepackage{MnSymbol}
\usetikzlibrary{math} 

\definecolor{acolor}{rgb}{0.1,.6,0}

\definecolor{rcolor}{rgb}{0.9,0.1,0.1}

\definecolor{bcolor}{rgb}{0.1,0,1}

\definecolor{dcolor}{rgb}{0.8,.1,.6}

\definecolor{mcolor}{rgb}{.9,.5,0.5}

\newcommand*{\cyl}{\vcenter{\hbox{\includegraphics[width=1.8em]{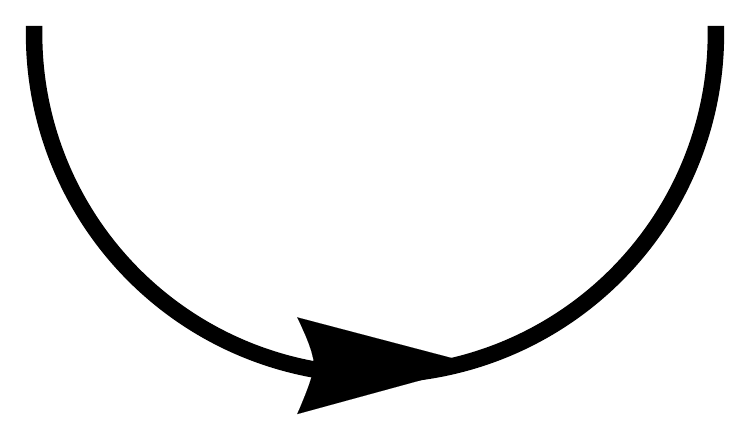}}}}

\title{On the nature of ensembles from gravitational path integrals}

\author{Donald Marolf} \emailAdd{marolf@ucsb.edu}

\affiliation{Department of Physics, University of California, Santa Barbara, CA 93106, USA}

\abstract{Spacetime wormholes in gravitational path integrals have long been known to lead to interpretations in terms of an ensemble of theories.  Here we probe the question of precisely what sort of theories such ensembles might contain.  Careful consideration of a simple $d=2$ topological model indicates that the Hilbert space structure of a general ensemble element fails to factorize over disconnected Cauchy-surface boundaries, and in particular that its Hilbert space ${\cal H}_{N_{CS\partial}}$ for $N_{CS\partial}$ Cauchy-surface boundaries fails to be positive definite when the number $N_{CS\partial}$ of disconnected such boundaries is large.  This observation then suggests a generalization of the AdS/CFT correspondence in which a bulk gravitational theory is dual to an ensemble of theories that deviate from standard CFTs by violating both locality and positivity (at least under certain circumstances).
Since violations of positivity are undesirable, we propose that positivity-violating elements of the ensemble be removed when studying physics in asymptotically AdS spacetimes (or in other contexts in which Cauchy surfaces have asymptotic boundaries), perhaps reducing the ensemble to a single standard CFT.  Nevertheless, properties of any remaining CFTs that are uncorrelated with positivity of ${\cal H}_{N_{CS\partial}}$ at large $N_{CS\partial}$will agree with those of typical elements of the full ensemble and may be computed using the ensemble average.

On the other hand, elements that violate positivity at large $N_{CS\partial}$  can still have a positive-definite  cosmological sector with $N_{CS\partial}=0$. The collection of such elements then defines a basis for a Hilbert space describing such cosmologies.  In contrast to the cases in which Cauchy-surfaces are allowed to have boundaries, we argue that the resulting Hilbert space need not decohere into single-state theories. As a result, familiar physics might be more easily recovered from this new scenario. }

\begin{document}

\maketitle

\section{Introduction}
\label{sec:intro}

Smooth manifolds in which disconnected components of the boundary are connected through the bulk are known as spacetime wormholes.
The inclusion of such wormholes in gravitational path integrals is typically discussed in Euclidean signature and has long been understood to encode non-local correlations that describe ensemble-like properties.  Work in the 1980's focused on tiny Planck-scale wormholes that led to an ensemble of local bulk couplings \cite{Coleman:1988cy,Giddings:1988cx,Giddings:1989ny}.  More recent works include effects from large wormholes which are generally less local, but which can still be interpreted in terms of an ensemble of theories \cite{Saad:2019lba,Penington:2019kki,Marolf:2020xie}.  For example, in contexts where gravitational path integrals might be thought to compute the inner product of two states, the presence of wormholes is naturally interpreted as implying that one is in fact considering an ensemble of theories, and that the value of the desired inner product depends on the particular member of the ensemble that one selects.  In particular, the simplest associated wormholes then compute the variance of the desired inner product across the ensemble.

This discussion is of particular interest in the context of the AdS/CFT conjecture.  The traditional form of this conjecture postulates that a single conformal field theory (CFT) is completely equivalent (or `dual') to string theory with appropriate asymptotically locally anti-de Sitter (AlAdS) boundary conditions \cite{Maldacena:1997re,Witten:1998qj}.  However, the fact that string theory contains gravity, and that it should thus in some sense include effects of spacetime wormholes, then suggests a modification of this conjecture in which the bulk string theory is in fact dual to an ensemble of CFTs (see e.g. \cite{Maldacena:2004rf}).   This suggestion is then supported by analyses of two-dimensional Jackiw-Teitelboim (JT) gravity which can, in a certain sense, be directly shown to be dual to an ensemble of theories \cite{Saad:2019lba}.  On the other hand, a problem with this suggestion is that -- at least in higher dimensions and in contexts with sufficient supersymmetry -- the number of CFTs appears to be too small to form a useful dual ensemble; see also additional comments in \cite{McNamara:2020uza,Schlenker:2022dyo}.

While one can imagine additional UV contributions that cancel or forbid the above wormhole contributions to the path integral  (see e.g. (\cite{Eberhardt:2020bgq,Eberhardt:2021jvj,Saad:2021rcu,Benini:2022hzx}), relying on such novel contributions raises the question of to which other computations they will also provide important corrections.  In particular, one wonders whether the Page curve calculations of \cite{Penington:2019npb,Almheiri:2019psf}, or more generally any computation of gravitational entropy, would be modified in significant ways.

A closely related issue in that (as we review below) the dual ensemble is expected to be defined by a preferred basis of what has been called `baby universe sector' of the bulk Hilbert space \cite{Coleman:1988cy,Giddings:1988cx,Marolf:2020xie}.  This is the sector of the theory that describes cosmological spacetimes which, in Lorentz signature, have compact Cauchy surfaces (whether they be a collection of spheres, tori, compact hyperbolic spaces, or more general manifolds).  Since we are interested in discussing such universes regardless of whether there is any `parent' universe (in which Cauchy surfaces might have boundaries), we will instead use the term Cauchy-surface-compact cosmologies (CSC-cosmologies, or CSCCs) to refer to such spacetimes below.  Other sectors of the theory will then describe settings in which Cauchy surfaces have non-trivial boundaries, in which we say that there are non-trivial Cauchy-surface boundaries (CS-boundaries).

A further point that we wish to emphasize is that,  as a result of the above connection, scenarios in which the dual ensemble degenerates to a single theory are expected to define a one-dimensional Hilbert space for such CSCCs.   One then naturally expects this to result in large discrepancies from naive semiclassical descriptions that allow an infinite dimensional space of such states; see e.g. comments in
\cite{Penington:2019npb,Almheiri:2019hni,Marolf:2020xie,Marolf:2020rpm,Usatyuk:2022afj}. Indeed, even if a large dual ensemble should happen to exist (so that the dimension of the CSCC Hilbert space is large),
in a state that describes spacetimes with Cauchy surfaces that contain both compact and non-compact components,
any process  acting at the CS-boundary that is sensitive to the differences between particular elements of the ensemble will generate entanglement with the CSC Hilbert space.  This entanglement will then permanently decohere the CSC Hilbert space space into the basis of preferred states that define the ensemble (the so-called $\alpha$-states).

\begin{figure}[h!]
        \centering
\includegraphics[width=0.5\linewidth]{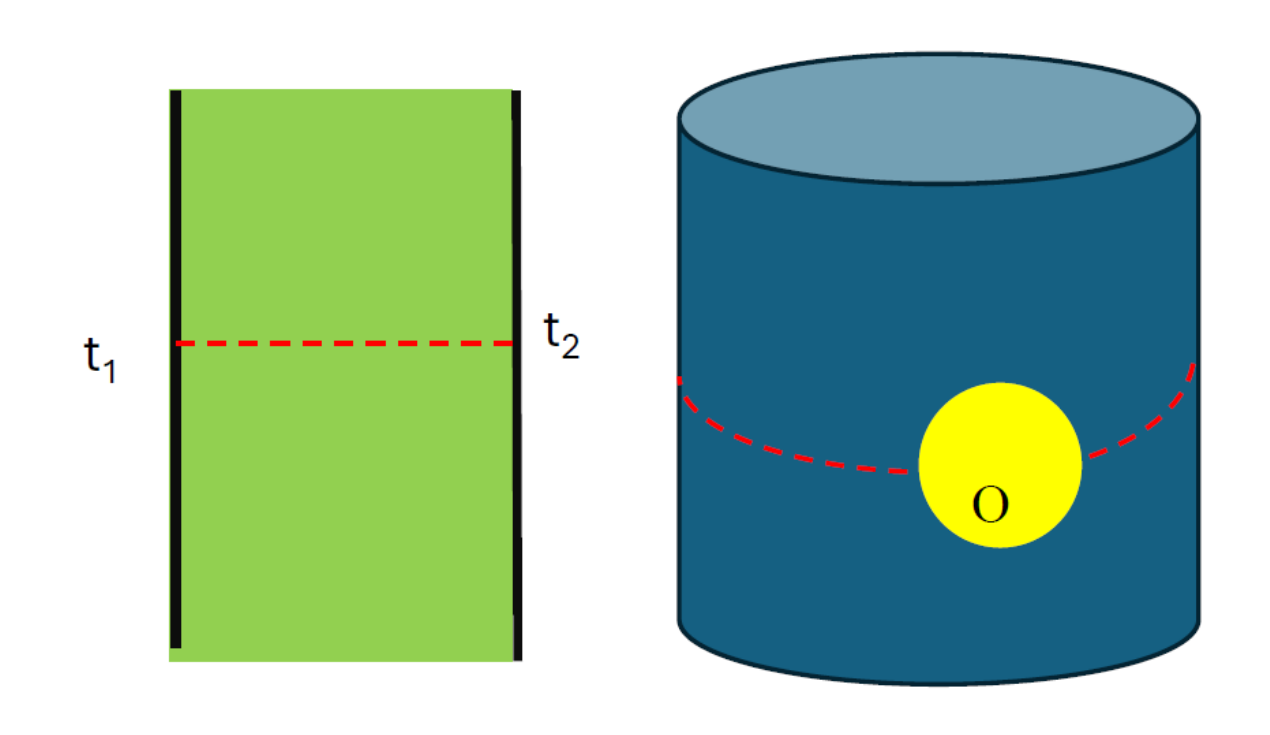}\caption{Observables that are not diagaonal in $\alpha$ will not commute with observables at Cauchy-surface boundaries (CS-boundaries).  A general such observable will thus fail to commute with the Hamiltonian at any CS-boundary and will thus depend on choices of times at all CS-boundaries, even in disconnected components of the Cauchy surface.  In particular, an observable $O$ that might be hoped to describe physics in a region (yellow) of a CSC-cosmology (blue), will nevertheless depend on choices of times $t_1,t_2$ at boundaries of disconnected components of spacetime (green).  Operations performed at such boundaries will thus tend to decohere physics described by $O$.}
\label{fig:timelabels}
\end{figure}

A related comment is that observables that are not diagaonal in $\alpha$ will generally fail to commute with observables at CS-boundaries, and in particular with the Hamiltonian at any such boundary, and so must be labeled by a `CS-boundary-time-function' that assigns a time coordinate to each point of the CS-boundary $\Sigma$.  This must be the case even for observables one hopes to use to describe physics in component of the Cauchy surface that are disconnected from all CS-boundaries; see figure \ref{fig:timelabels}.   It is this time-dependence that generally leads to the above decoherence.   Thus, even if there is a large ensemble, one seems to find large discrepancies from naive semiclassical predictions.  We will refer to this issue below as the cosmological decoherence problem\footnote{\label{foot:dec} A related mathematical problem is that observables defined by inserting asymptotic boundaries (say, at Euclidean past and future) into the path integral are superselected regardless of the presence of non-trivial CSC-boundaries.  Indeed, this is how the $\alpha$ states are defined in the framework of \cite{Marolf:2020xie}.  So, even in the absence of the above decohering processes, a new mathematical construction would be required to describe any physics associated with coherent superpositions of $\alpha$-states.}.

The general set of issues just described is collectively known as the AdS/CFT factorization problem.  In an attempt to find new scenarios which could lead to a more satisfactory resolution of this problem, we will ask in this work whether a gravitational description can be dual to an ensemble of theories that are somehow more general than CFTs (and presumeably also more general than standard local quantum fields theories).  This is an idea that has also been explored in the context of AdS$_3$/CFT$_2$ for the purely practical reason that general CFT$_2$'s are hard to enumerate \cite{Maloney:2020nni,Datta:2021ftn,Chandra:2022bqq,Belin:2023efa}.  We will first investigate this question at a purely mathematical level, saving discussions of the physical implications of having such more general theories in the dual ensemble for the discussion in section \ref{sec:disc}.  This is in part because it is difficult to address such physical implications without first having some idea of the particular manner in which such more general theories fail to be CFTs.

Our approach to this question will be to recall that, at least in some cases, the gravitational path integral can be use to {\it directly} construct the relevant ensemble; see e.g. the discussion in \cite{Marolf:2020xie}.  As mentioned above, this construction involves the detailed structure of the part of the Hilbert space that describes CSC-cosmologies.   However, for anything approaching a realistic gravity theory, the construction is handicapped by the fact that the path integral is generally not understood beyond the semiclassical approximation.  Indeed, for a path integral defined by a sum over geometries and topologies, the sum over topologies may fundamentally fail to converge.  As a result, it has meaning only as an asymptotic series in quantities that are exponentially small in $1/G_N$ when $G_N$ is small; see e.g. \cite{Saad:2019lba}.

It is therefore useful to turn to simple models in which we can explicitly construct the baby universe sector and, in particular, the basis of $\alpha$-states whose members define the ensemble.
Below, we return to the $d=2$ topological gravity model of \cite{Marolf:2020xie}.  This model depends on two couplings $S_0$ and $S_\partial$.  While at some level the model was fully solved in
\cite{Marolf:2020xie}, that work emphasized the case $S_0=S_\partial$ for which the results are particularly simple and for which the structure most resembles an ensemble of standard quantum theories.  In contrast, we will now describe consequences of the analysis of \cite{Marolf:2020xie} for the more general case $S_0\neq S_\partial$ in an effort to gain insight into how bulk systems might be dual to an ensemble (or to a suitable generalization thereof) of more general theories.  We will also explore the dependence of the properties of $\alpha$-states on $S_0,S_\partial$.

We begin in section \ref{sec:framework} with a review the basic analysis of \cite{Marolf:2020xie} for the construction a Hilbert space from the path integral. As mentioned above, this leads to an ensemble of superselection sectors associated with a preferred basis (the $\alpha$-state basis) for the Hilbert space of Cauchy-surface-compact cosmologies. We also take this opportunity to make some clarifications, to add further comments, and in particular to rephrase the discussion of Hilbert space sectors with asymptotic boundaries.

The specific topological model studied in detail by \cite{Marolf:2020xie} is then reviewed in section \ref{sec:results} for the case without end-of-the-world branes, where we now more carefully examine the implications for general values of the two couplings $S_0, S_\partial$ in light of more recent results from \cite{Colafranceschi:2023urj}.  The case with
end-of-the-world branes is deferred to appendix \ref{subsec:withEOW}, where we also introduce a third coupling $S_{EOW}$ that was not previously studied.  The stage is then set for the discussion in
section \ref{sec:disc} where we attempt to draw lessons for more general (and perhaps higher dimensional) models of UV-completions of quantum gravity theories.   In doing so, we will describe how such lessons might resolve the AdS/CFT factorization problem in general, and the associated resolution of the cosmological decoherence problem in particular.

\section{Quantum gravity, baby universes, and $\alpha$-states: a framework}
\label{sec:framework}

We now briefly review the framework described in \cite{Marolf:2020xie} for understanding the space of quantum gravity states and its relation to baby universe states.  In doing so, we consider a theory defined by an object that we may call a Euclidean bulk path integral $\zeta$.  We will think of this $\zeta$ as providing a UV completion of a bulk gravitational theory, and we will require $\zeta$ to have certain properties that appear natural in this context.  However, consistent with the fact that we do not constrain the form of the above UV completion, we will {\it not} require $\zeta$ to be defined by a sum over metrics, or  to be fundamentally defined as an integral of any form.  However, in cases where $\zeta$ is in fact defined by such an integral, we imagine that we can evaluate the integral for various allowed boundary conditions $J$.  In this context, we may write

\begin{equation}
\label{eq:EPI}
    \zeta[J] :=\int_{\Phi \sim J} \mathcal{D} \Phi e^{-S[\Phi]},
\end{equation}
where the notation $\Phi \sim J$ indicates that the integral is restricted to field configurations $\Phi$ that satisfy the given boundary condition $J$.

Regardless of whether $\zeta$ takes the form \eqref{eq:EPI}, we will assume that $\zeta$ defines a function on a class $\mathcal J$ of objects $J$ that we may call boundary conditions.  It is often useful to think of our boundary conditions as being asymptotically locally anti-de Sitter (AlAdS), though we will not necessarily require this to be the case.  We will, however,  assume that a given boundary condition $J$ specifies a manifold $M[J]$ (`the boundary manifold') along with some class of local fields\footnote{The generalization to allowing non-local fields is straightforward, but naturally leads to mixed states.  It thus requires a discussion in terms of density matrices.  For simplicity of the current presentation we restrict to local boundary conditions in order to use the pure-state formalism that may be more transparent to many readers.} (also called {\it sources}) $\Phi_\partial[J]$ on $M$.  We will also take $\zeta$ to be invariant under diffeomorphisms of any $M[J]$.  In a simple context one might expect a $D$-dimensional bulk quantum gravity theory to have boundary manifolds $M[J]$ of dimension $d=D-1$.  Nonetheless, when the bulk spacetimes have compact dimensions that approach a constant size at infinity the useful notion of `boundary manifold' $M[J]$ may well have dimension $d$ with $D-d > 1$. Finally, it would also be natural to require the manifolds $M[J]$ to be compact and closed (with $\partial M[J]=\emptyset$), but we will not do so explicitly.  Theories in which $\partial M[J]$ is allowed to be non-trivial can be interpreted as including end-of-the-world (EOW) branes.

The class of boundary conditions $\mathcal J$ will be required to admit an involution $*$.  In particular,
For every $J \in \mathcal J$ we assume there to be some $J^*\in \mathcal J$ with $M[J^*] = M[J]$ such that $J^{**}=J$ and
\begin{equation}
\label{eq:conj}
\zeta[J^*] = \left(\zeta[J] \right)^*.
\end{equation}
Since $\zeta[J]$ is just a complex number, on the right-hand size of \eqref{eq:conj} the operation ${}^*$ is standard complex conjugation.  But we have not specified the precise way that $*$ acts on general boundary conditions $J$.  In general, this action will be a form of CPT-conjugation rather than just complex-conjugation of sources.
We will refer to $J^*$ as the conjugate boundary condition to $J$, or as just the conjugate of $J$ for short.

When the manifold $M[J]$ is a disjoint union of $n$ connected components $M_i$, we will write $J = J_1 \sqcup J_2 \sqcup \dots \sqcup J_n$ where $M_i = M[J_i]$ and where the sources $\Phi_\partial[J_i]$ are the restrictions to $M_i$ of $\Phi_{\partial}[J]$.
Motivated by the special case \eqref{eq:EPI}, we will assume the functional $\zeta$ and the set of boundary conditions $\mathcal J$ to have four further important properties.
The first is that $\mathcal J$ includes the trivial boundary condition described by the empty set $J=\emptyset$, so that $\zeta[\emptyset]$ is well-defined.  The second is that $\mathcal J$ is closed under disjoint unions; i.e., for $J_1,J_2 \in \mathcal J$ we must also have $J_1 \sqcup J_2 \in \mathcal J$. The third is that the operation ${}^*$ preserves disjoint unions in the sense that we have
\begin{equation}
\left(J_1 \sqcup J_2 \right)^* = J_1^* \sqcup J_2^*.
\end{equation}
Finally, the fourth is that for $J = J_1 \sqcup J_2 \sqcup \dots \sqcup J_n$ the path integral $\zeta[J]$ does not depend on the ordering assigned to the various connected components of the boundary.  Thus $\zeta[J_1 \sqcup J_2 \sqcup \dots \sqcup J_n]$ is invariant under permutations of the $J_i$;  e.g.
\begin{equation}
\label{eq:perminv}
  \zeta\left[ J_1 \sqcup J_2 \sqcup J_3\right] =     \zeta\left[ J_1 \sqcup J_3 \sqcup J_2\right] =
    \zeta\left[ J_3 \sqcup J_2 \sqcup J_1\right], \  {\rm  etc.}
\end{equation}
However, since the special case \eqref{eq:EPI} would naturally allow contributions from spacetime wormholes, we expect that $\zeta$ will generally fail to factorize over such disjoint unions; i.e., in general we have
\begin{equation}
  \zeta\left[ J_1 \sqcup J_2 \right] \neq    \zeta\left[ J_1\right] \zeta\left[ J_2\right].
\end{equation}

\subsection{The Cauchy-surface-compact cosmological sector (aka the baby universe sector)}
\label{subsec:BU}

We now wish to use $\zeta$ to define a space of states, as well as an inner product on that space.  Since we think of $\zeta$ as providing a UV completion for some theory of quantum gravity we will use the symbol ${\mathcal H}_{QG}$ to denote the final space of states so obtained.  However, we forewarn the reader that ${\mathcal H}_{QG}$ may not always be a Hilbert space. Following \cite{Marolf:2020xie}, we begin with the baby universe sector of the space of states. This sector ${\cal H}_{CSCC}$ may equivalently be called the sector of Cauchy-surface-compact cosmologies (CSCCs), or the sector in which Cauchy surfaces have no boundaries (i.e., the sector without CS-boundaries), though such Cauchy surfaces may have multiple connected components (e.g., the Cauchy surface might be a set of disconnected spheres).  In a description in terms of spacetimes, we then have in mind that such spacetimes may have boundaries (and must thus respect boundary conditions) in the asymptotic past or future (in either a Euclidean or a Lorentzian sense), but that (in some Lorentzian description) the Cauchy surfaces for such spacetimes would be compact manifolds-without-boundary.

 For each boundary condition $J \in \mathcal J$, we will define a state $|J\rangle \in {\cal H}_{CSCC}$, and we take the path integral to define an inner product
\begin{equation}
\label{eq:JJpip}
\langle J|J'\rangle = \zeta[J^*\sqcup J']
\end{equation}
for all $J, J'\in \mathcal J$.  It is then straightforward to extend the inner product \eqref{eq:JJpip} to the space ${\mathds C}^{\mathcal J}$ of formal finite linear combinations $\sum_{i=1}^n c_i |J_i\rangle$ of states $|J_i \rangle$ for $N\in {\mathbb Z}^+$ and $c_i \in {\mathbb C}$. To do so, we simply take the inner product to be anti-linear in the first argument and linear in the second. We emphasize that this is a {\it new} formal linear structure and that it is unrelated to any linear structure on the space of sources.  In particular, given boundary conditions $J_1,J_2$ which define the same boundary manifold $M[J_1]=M[J_2]$ and sources $\Phi_\partial[J_1], \Phi_\partial[J_2]$, the state $|J_1 \rangle + |J_2\rangle$ generally has nothing to do with any boundary condition $J_3$ that might in any sense have $M[J_3] = M[J_1]=M[J_2]$ and $\Phi_\partial[J_3] = \Phi_\partial[J_1] + \Phi_\partial[J_2]$.

The above definition will generally result in a non-trivial null space ${\cal N} \subset {\mathcal C}[\mathcal J]$ such that, for $|\nu\rangle \in {\cal N}$ and all $|\psi\rangle \in {\mathds C}^{\mathcal J}$, we have
\begin{equation}
\langle \psi|\nu \rangle =0.
\end{equation}


Due to the presence of the null space $\mathcal N$, it is natural to pass to the quotient space ${\mathds C}^{\mathcal J}/ \mathcal N$, on which the induced inner product is non-degenerate.
Note that the dimension of ${\mathds C}^{\mathcal J}/ \mathcal N$ will tend to be infinite even in extremely simple toy models due to the fact that $\mathcal J$ includes boundary conditions for which $M[J]$ has arbitrarily many connected components.  I.e., for any $J \in \mathcal J$, we will also have boundary conditions $J\sqcup J$, $J\sqcup J \sqcup J$, etc. which will generally define linearly independent states in ${\mathds C}^{\mathcal J}/ \mathcal N$.

When the resulting inner product on ${\mathds C}^{\mathcal J}/ \mathcal N$ is positive-definite (and thus Hermitian), we define the Hilbert space of Cauchy-surface-compact cosmologies ${\mathcal H}_{CSCC}$ to be the associated completion of
${\mathds C}^{\mathcal J}/ \mathcal N$ defined by first constructing all Cauchy sequences in ${\mathds C}^{\mathcal J}/ \mathcal N$  and then again taking the quotient by the set of null states.
We will assume this positivity throughout the rest of this section, and we will henceforth use the notation $|J\rangle$ to describe the state on ${\mathcal H}_{CSCC}$ defined  by $J\in \mathcal J$.  Note that the permutation-invariance \eqref{eq:perminv} then implies the states $|J_1 \sqcup J_2 \sqcup \dots \sqcup J_n\rangle$ to be invariant under permutations of the $J_i$; i.e.,
\begin{equation}
\label{eq:stateperminv}
|J_1 \sqcup J_2 \sqcup J_3 \rangle = |J_2 \sqcup J_1 \sqcup J_3 \rangle = |J_2 \sqcup J_3 \sqcup J_1 \rangle, \ \ \ {\rm etc}.
\end{equation}
The construction of $\mathcal H_{CSCC}$ from Cauchy sequences also implies that ${\mathds C}^{\mathcal J}/ \mathcal N$  can be regarded as a dense subspace of $\mathcal H_{CSCC}$.  In this context it is useful to refer to ${\mathds C}^{\mathcal J}/ \mathcal N$ as the dense domain ${\mathcal D}_{\mathcal J}$.

The ensemble description arises from the observation that any $J \in \mathcal J$ also induces a densely-defined operator $\widehat {Z[J]}$ on $\mathcal H_{CSCC}$.  In particular, for any $J'\in \mathcal J$ we simply define
\begin{equation}
\label{eq:ZJop}
\widehat {Z[J]} |J'\rangle : = |J \sqcup J'\rangle,
\end{equation}
which is manifestly linear in $J'$ as desired.  The states \eqref{eq:ZJop} are all of finite norm but, as we will see explicitly in the topological model below, the operators $\widehat {Z[J]}$ are generally unbounded as this norm can diverge when taking limits that converge in $\mathcal H_{CSCC}$ but not in ${\cal D_{J}}$.  As a side comment, we also note that the original path integral can now be understood as the expectation value functional of such operators in the `no-boundary' state $|\emptyset\rangle$:
\begin{equation}
\label{eq:zetaNB}
\zeta[J] = \langle \emptyset | \widehat {Z[J]} |\emptyset \rangle.
\end{equation}

It is useful to recall that for any $J, J', J''\in \mathcal J$ we have
\begin{equation}
\label{eq:Zadj1}
\langle J''| \widehat {Z[J]} |J'\rangle : = \langle J'' |J \sqcup J'\rangle =\zeta[J''{}^* \sqcup  J \sqcup J'] = \langle J''  \sqcup J^*| J'\rangle .
\end{equation}
As a result, when acting on the dense domain ${\cal D_{J}}$ we have
\begin{equation}
\label{eq:Zadj2}
\left(\widehat {Z[J]}\right)^\dagger = \widehat {Z[J^*]}.
\end{equation}
However, as noted above,  these operators are generally unbounded.  As a result, detailed analysis in the context of particular models will likely be required to determine if \eqref{eq:Zadj2} holds more generally, or if can be made to hold for all $J\in \mathcal J$ by choosing appropriate extensions of the above $\widehat{Z[J]}$.  We will nevertheless follow \cite{Marolf:2020xie} in assuming that such an extension is both possible and unique.  In that case, it is natural to again denote the extension by $\widehat{Z[J]}$.  Note that $\widehat{Z[J]}_{\pm} := \widehat{Z[J]} \pm \widehat{Z[J^*]}$ are then self-adjoint (or anti self-adjoint).

Let us also observe that the permutation invariance \eqref{eq:stateperminv} implies commutativity of two operators $\widehat Z[J], \widehat Z[J']$ when acting on ${\cal D_{J}}$; i.e., on ${\cal D_{J}}$ we have
\begin{equation}
\label{eq:Zcom}
\widehat {Z[J]} \widehat {Z[J']} = \widehat{Z[J\sqcup J']} = \widehat {Z[J']} \widehat {Z[J]} \ \ \ \forall J,J' \in \mathcal J.
\end{equation}
Again, since such operators can be unbounded, a detailed analysis is generally required to determine if \eqref{eq:Zcom} holds on states outside   ${\cal D_{J}}$.  See e.g. \cite{RS} for an example where there are no extensions that satisfy both \eqref{eq:Zadj2} and \eqref{eq:Zcom}.  Nonetheless, we will again follow \cite{Marolf:2020xie} in assuming that the extended $\widehat{Z[J]}$ satisfying \eqref{eq:Zadj2} continue to respect the condition \eqref{eq:Zcom} in the sense that bounded functions of $\widehat{Z[J]}_{\pm}$ commute with bounded functions of $\widehat{Z[J']}_{\pm}$ on the entire space $\mathcal H_{CSCC}$ of CSC-cosmology  states\footnote{After thus extending the $\widehat{Z[J]}$ for connected $M[J]$, we can then enforce the rest of \eqref{eq:Zcom} by defining   $\widehat{Z[J]}$ for $J = J_1 \sqcup J_2$ by the relation $\widehat{Z[J]}:= \widehat{Z[J_1]} \widehat{Z[J_2]}$.}.

In particular, taking $J'=J^*$ we see that  $\widehat{Z[J]}$ commutes with $\widehat{Z[J]}^\dagger$, or equivalently that $\widehat{Z[J]}_{+}$ commutes with $\widehat{Z[J]}_{-}$.  We can thus diagonalize each $\widehat{Z[J]}$, and the above extension of \eqref{eq:Zcom} tells us that we can do so simultaneously for all $J \in \mathcal J$.  We will denote the joint spectrum of all $\widehat{Z[J]}$ by $\mathcal A$. For $\alpha \in \mathcal A$, the corresponding simultaneous eigenstate will be called $|\alpha \rangle$.    Such $|\alpha \rangle$ are the analogues in this formalism of the $\alpha$-states of \cite{Coleman:1988cy,Giddings:1988cx}.

The eigenvalue of  $\widehat{Z[J]}$ in the state $|\alpha \rangle$ will be written $Z_\alpha[J]$, so that
\begin{equation}
\label{eq:ZJalpha}
\widehat Z[J]  |\alpha \rangle = Z_\alpha[J] |\alpha \rangle .
\end{equation}
Distinct simultaneous eigenstates are of course orthogonal, so that $\langle \alpha |\alpha'\rangle =0$ for $\alpha \neq \alpha'$.  They are also complete.  Thus for some measure $d\alpha$ we have
\begin{equation}
\label{eq:idalpha}
{\mathds 1} = \int d\alpha  |\alpha \rangle \langle \alpha |,
\end{equation}
and thus
\begin{equation}
\label{eq:id2alpha}
|\alpha' \rangle \langle \alpha' | = {\mathds 1} |\alpha' \rangle \langle \alpha' |= \int d\alpha   |\alpha \rangle \langle \alpha |\alpha' \rangle \langle \alpha' |.
\end{equation}
When the joint spectrum $\mathcal A$ is continuous, this means that $\langle \alpha |\alpha' \rangle$ is a Dirac delta-function with respect to the measure $d\alpha$.

In analogy with \eqref{eq:zetaNB}, we can now define the `path integral in a given $\alpha$-state' to be the functional
\begin{equation}
\label{eq:zetaalpha}
\zeta_\alpha[J] := \langle \alpha | \widehat {Z[J]} |\alpha \rangle = Z_\alpha[J].
\end{equation}
We will shortly explain why it is natural to call \eqref{eq:zetaalpha} a `path  integral.'   First, however, let us
note that, since we assumed \eqref{eq:Zcom} to hold on the entire Hilbert space, we also have $Z_\alpha[J_1 \sqcup J_2] = Z_\alpha[J_1] Z_\alpha[J_2].$  As a result, the `path integral' $\zeta_\alpha$ factorizes over disconnected boundaries. The lack of factorization of the original path integral $\zeta$ is then understood as resulting from the fact that
\eqref{eq:idalpha} requires the no-boundary state to take the form
\begin{equation}
\label{eq:NBalpha}
|\emptyset\rangle = \int d\alpha  |\alpha \rangle\langle \alpha |\emptyset \rangle.
\end{equation}
Thus the original path integral $\zeta$ may be written in the form
\begin{eqnarray}
\label{eq:zetaalphadecomp}
\zeta[J] &=& \langle \emptyset | \widehat { Z[J]} |\emptyset \rangle \cr &=& \int d\alpha \langle \emptyset  |\alpha \rangle \langle \alpha | \widehat { Z[J]} |\emptyset \rangle \cr
&=& \int d\alpha \zeta_{\alpha}[J] \Big| \langle \alpha|\emptyset \rangle \Big|^2;
 \end{eqnarray}
i.e., the original path integral is an average over the ensemble $\mathcal A$ of the $\alpha$-state path integrals $\zeta_\alpha$ weighted by the measure $\Big| \langle \alpha|\emptyset \rangle \Big|^2 d \alpha$.  While each $\zeta_\alpha$ factorizes over disconnected boundaries, the ensemble average in \eqref{eq:zetaalphadecomp} imposes correlations that prevent the corresponding factorization of the original $\zeta$.

To explain why \eqref{eq:zetaalpha} can be called a path integral, let us recall from the general theory of von Neumann algebras that the projector $|\alpha \rangle \langle \alpha |$ can always be described as a limit of sums of products of the $\widehat{ Z[J_i]}$.  In this sense it is a (perhaps distributional) function of these operators.  As such, it makes sense to insert $|\alpha \rangle \langle \alpha|$ into our path integral along with any $\widehat{ Z[J_i]}$ and to write
\begin{eqnarray}
\label{eq:zetaalpha1_5}
\zeta\left[\widehat {Z[J]} |\alpha \rangle\langle \alpha |\right] =
 \langle \emptyset |\widehat {Z[J]} |\alpha \rangle\langle \alpha |\emptyset \rangle].
\end{eqnarray}
Indeed, the right-hand side is continuous under the above-mentioned limit and so can be used to define the left-hand side.  Normalizing the result then gives
\begin{eqnarray}
\label{eq:zetaalpha2}
  \frac{\zeta\left[\widehat {Z[J]} |\alpha \rangle\langle \alpha |\right]}{\zeta\left[|\alpha \rangle\langle \alpha |\right]}=
 \frac{\langle \emptyset |\widehat {Z[J]} |\alpha \rangle\langle \alpha |\emptyset \rangle]}{\langle \emptyset | |\alpha \rangle\langle \alpha |\emptyset \rangle}
 = \langle \alpha | \widehat {Z[J]} |\alpha \rangle = Z_\alpha[J] = \zeta_\alpha[J].
\end{eqnarray}
Thus $\zeta_\alpha$ is simply a renormalized version of the original path integral $\zeta$ with an extra insertion of $|\alpha \rangle\langle \alpha |$.     The factorizing path integral \eqref{eq:zetaalpha} is thus built from $\zeta$ by inserting (limits of linear combinations of) extra boundaries.  We note that,  since each $\widehat {Z[J]}$ commutes with
$|\alpha \rangle\langle \alpha |$, there is no need to insert these boundaries in any particular order.

If one wishes, one can now perform various resummations of the contributions to the path integral generated by the extra boundaries introduced by $|\alpha \rangle\langle \alpha |$.  One option is to use an appropriate `bulk-to-boundary propagator' to move these contributions into the bulk and to call the result a `factorizing brane' of the sort described in e.g. \cite{Blommaert:2021fob}.  Another is to think of wormholes connecting any bulk spacetime with the new boundaries introduced by $|\alpha \rangle\langle \alpha |$ as new dynamical objects to be introduced into the theory.  These objects are naturally called `half-wormholes' as they have only one end in the bulk spacetime (since the other end effectively ends on the `factorizing brane' described above).  The path integral \eqref{eq:zetaalpha} may then be rewritten without explicit reference to the projection $|\alpha \rangle\langle \alpha |$ and the boundaries it inserts by instead stating appropriate rules for summing over contributions with half-wormholes as in \cite{Saad:2021rcu,Saad:2021uzi}.  The CSCC $\alpha$-state formalism thus provides a convenient language for describing the relations between all of these different formulations and seeing that they are fundamentally equivalent.   The descriptions of these connections has been made especially explicit in \cite{Blommaert:2022ucs} for Jackiw-Teitelboim gravity and in \cite{Marolf:2020xie} for the topological model we study below.

An important property of the present formalism is that there can be only a single $\alpha$-state with any given set of eigenvalues $Z_\alpha[J]$.  This follows from the fact that, up to an overall constant factor $\langle \emptyset | \alpha\rangle$, the inner products $\langle J | \alpha\rangle$ are all determined by the eigenvalues via
\begin{equation}
\langle J | \alpha\rangle = \langle \emptyset |\widehat{Z[J]} | \alpha\rangle = Z_\alpha[J] \langle \emptyset | \alpha\rangle.
\end{equation}
Since the states $|J\rangle$ span a dense subspace $\mathcal D_{\mathcal J}$ of $\mathcal H_{CSCC}$, the inner products $\langle J | \alpha\rangle$ then uniquely determine the state $|\alpha\rangle$.  This in particular means that $|\alpha \rangle\langle \alpha |$ is a {\it one-dimensional} projection.  As a result, after inserting $|\alpha \rangle\langle \alpha |$ into a path integral, any further insertions that one makes can serve only to renormalize the path integral (potentially by the degenerate factor zero) and cannot otherwise change the associated CSCC state.

In particular, any $\alpha$-sector defines a unique state in the Hilbert space $\mathcal H_{CSCC}$ describing Cauchy-surface-compact cosmologies.  This is true even in cases where the ensemble defined by the above no boundary state turns out to be trivial in the sense that it contains only a single $\alpha$-state.  Indeed, as illustrated by the above discussion, such a case is indistinguishable from what one obtains by starting with a path integral $\zeta$ that defines a larger ensemble and then inserting the projection $|\alpha \rangle\langle \alpha |$ to construct $\zeta_\alpha$.  In such contexts, up to normalization (including a possible phase) any two non-trivial states of $\mathcal H_{CSCC}$ can differ only by a null state. In particular, as recently reemphasized in  \cite{Usatyuk:2024mzs},  all states in  $\mathcal H_{CSCC}$  are equivalent to the no-boundary state $|\emptyset\rangle$.  As noted in \cite{Marolf:2020xie}, this observation explains why island computations always find zero entanglement with Cauchy-surface-compact cosmologies \cite{Penington:2019npb,Almheiri:2019hni}.  Of course, this feature is also in strong tension with the usual way we understand physics in closed universes.   However, we will postpone discussion of possible resolutions to section \ref{sec:disc}, where we can take inspiration from features of the model discussed in section \ref{sec:results} below.

\subsection{Sectors with non-trivial asymptotic boundaries at spatial infinity}
\label{subsec:Sigma}

While the CSC-cosmology sector $\mathcal H_{CSCC}$ is the closure of the space spanned by the states $|J\rangle $ for $J\in \mathcal J$, the local nature of our boundary conditions also allows the construction of other sectors associated with spacetimes whose Cauchy surfaces (in some Lorentzian setting) can have non-trivial asymptotic boundaries; e.g., in some asymptotically AdS region.   One way to set up this construction is to postulate that, for some set $\sigma[\mathcal J]$ of manifolds $\Sigma$ endowed with appropriate fields, there is a set of `half-sources' $\mathcal J_{\Sigma}$ with the following properties:
\begin{enumerate}[label=\roman*)]
\item{} Each $J_\Sigma \in \mathcal J_{\Sigma}$ is a $d$-manifold $M[J_\Sigma]$ with $(d-1)$-dimensional non-dynamical boundary\footnote{This terminology is intended to allow $\Sigma$ to be distinguished from e.g. end-of-the-world-brane boundaries of $M[J_\Sigma]$.  Such end-of-the-world-brane boundaries of $M[J_\Sigma]$ would then be classified as {\it dynamical} boundaries and would be treated as part of the `local field data' in a given boundary condition $J_\Sigma$, but would not be specified by the label $\Sigma$ alone.  Note that since the dimension $d-1$ of any boundary of $M[J_\Sigma]$ will be strictly less than $D-1$ (with $D$ the dimension of the bulk spacetime),  an end-of-the-world-brane boundary of $M[J_\Sigma]$ represents a source that inserts EOW-branes (or anti-branes) into a given spacetime whence they will propagate further in a manner not directly constrained by $J_\Sigma$.} $\Sigma$ together with a collection of local fields on $M[J_\Sigma]$.

\item{} For $J_\Sigma \in \mathcal J_{\Sigma}$, suppose that $M[J_\Sigma] = M^0_\Sigma \sqcup M_\emptyset$ where $M_\emptyset$ has no non-dynamical boundary.  Then the restriction of the boundary fields $\Phi_\partial[J_\Sigma]$ to $M_\emptyset$ defines a boundary condition $J_\emptyset \in \mathcal J$, and the restriction to $M^0_\Sigma$ defines a boundary condition $J^0_\Sigma \in \mathcal J_\Sigma$.

\item{} For each $J_\Sigma \in \mathcal J_{\Sigma}$, there is a pair $J^*_\Sigma = (M[J^*_\Sigma], \Phi_\partial[J^*])$ built from a manifold $M[J^*_\Sigma]$ with non-dynamical boundary  $\Sigma$ and a set of local fields $\Phi_\partial[J^*]$ on $M[J^*_\Sigma]$.  For all $J_{\Sigma}, J_{\Sigma}{}' \in  \mathcal J_{\Sigma}$ we require that we can define a boundary condition $J'{}_\Sigma^* \cup_\Sigma J_\Sigma \in \mathcal J$  by gluing together  $J'{}_\Sigma^*$ and $J_\Sigma$ across $\Sigma$; i.e., by taking the union of the two manifolds and taking the two copies of $\Sigma$ to be identified.  We also require this construction to satisfy
    \begin{equation}
    \left(J'{}_\Sigma^* \cup_\Sigma J_\Sigma\right)^* \cong J_\Sigma^* \cup_\Sigma J'_\Sigma,
    \end{equation}
where $\cong$ represents equivalence modulo diffeomorphisms.  Thus
    \begin{equation}
    \label{eq:zetaskew}
    \zeta\left[J'{}_\Sigma^* \cup_\Sigma J_\Sigma\right] = \zeta\left[ J_\Sigma^* \cup_\Sigma J'_\Sigma\right]^*.
    \end{equation}
In particular, $\zeta[J_\Sigma^* \cup_\Sigma J_\Sigma]$ is real.
\end{enumerate}
In this context, even in a Euclidean setting, we will refer to $\Sigma$ as the CS-boundary.  In a Lorentz-signature setting we would indeed expect all Cauchy surfaces for (semi-)classical spacetimes to end on $\Sigma$.  Note that CS-boundaries are of dimension $d-1$ while the boundaries $M$ of bulk spacetimes (which we call simply `boundaries') are of dimension $d$.

\begin{figure}[h!]
        \centering
\includegraphics[width=0.3\linewidth]{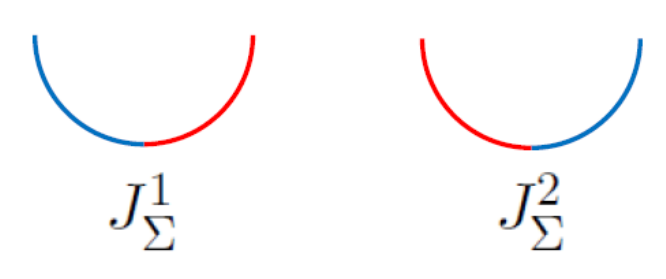}\hspace{2cm}\includegraphics[width=0.4\linewidth]{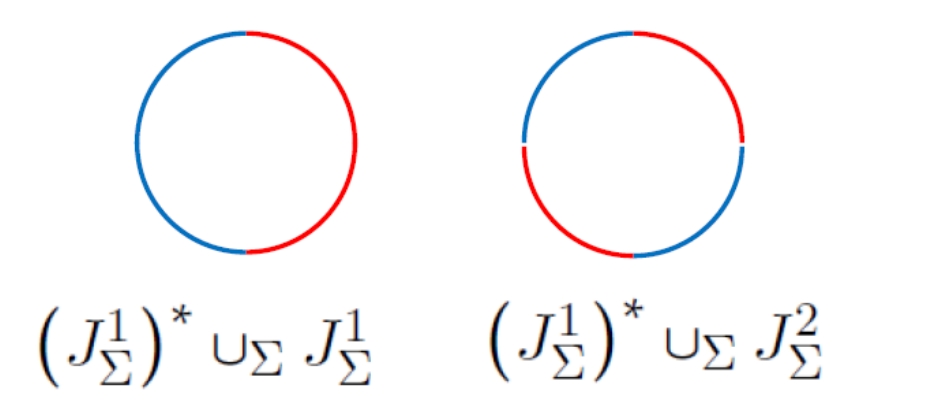}\caption{Simple cartoons of possible half-sources $J^1_\Sigma,J^2_\Sigma$ are shown at left as unoriented line segments with red/blue regions indicating distinct local source fields. Here $\Sigma$ is just a pair of points (without orientations).   Note that $J^1_\Sigma$ is related to $J^2_\Sigma$ by a diffeomorphism that acts as a right/left reflection.  Nevertheless, as shown at right, the complete source  $\left(J^1_\Sigma\right)^* \cup_\Sigma J^1_\Sigma$ is not diffeomorphic to the complete source
$\left(J^1_\Sigma\right)^* \cup_\Sigma J^2_\Sigma$.}
\label{fig:diffbndy}
\end{figure}

While the path integral $\zeta$ is invariant under diffeomorphisms, half-sources $J^1_\Sigma, J^2_\Sigma$ related by diffeomorphisms that act non-trivially on $\Sigma$ are generally {\it not} equivalent.  This is because $\left(J^1_\Sigma\right)^* \cup_\Sigma J^1_\Sigma$ and
$\left(J^1_\Sigma\right)^* \cup_\Sigma J^2_\Sigma$ are generally not diffeommorphic; see figure \ref{fig:diffbndy}.

The reality condition \eqref{eq:zetaskew}, enables us to use
the path integral $\zeta$ to define a skew-symmetric inner product on the vector space ${\mathds C}^{\mathcal J_\Sigma}$ of formal complex linear combinations of the states $|J\rangle$ for $J \in \mathcal J_\Sigma$.  In particular, we have
\begin{equation}
\langle J'_\Sigma| J_\Sigma \rangle = \zeta\left[J'{}_\Sigma^* \cup_\Sigma J_\Sigma\right].
\end{equation}
The inner product of more general linear combinations is then defined to be anti-linear in the first argument and linear in the second.  When this inner product is positive semi-definite, we can define a Hilbert space $\mathcal H_\Sigma$ by completing ${\mathds C}^{\mathcal J_\Sigma}$ through the addition of Cauchy sequences and taking the quotient of the the result by the set of null states.    We take the full quantum gravity Hilbert space $\mathcal H_{QG}$ to be the direct sum over sectors with all possible Cauchy-surface boundaries
\begin{equation}
\label{eq:HQG1}
\mathcal H_{QG} = \oplus_{\Sigma \in \sigma[\mathcal J]} \mathcal H_\Sigma.
\end{equation}

We now impose the following natural requirements on the set $\sigma[\mathcal J]$:
\begin{enumerate}[label=\arabic*)]
\item{} The set $\sigma[\mathcal J]$ is closed under disjoint union; i.e., for $\Sigma, \Sigma' \in \sigma[\mathcal J]$ we have $\Sigma \sqcup \Sigma' \in \sigma[\mathcal J]$.

\item{} The empty set $\emptyset$ is a member of $\sigma[\mathcal J]$.  Furthermore,  we take $\mathcal J_\emptyset := \mathcal J$. Conditions (i), (ii), and (iii) above are then manifestly satisfied by ${\mathcal J}_\emptyset$, and the resulting Hilbert space is just $\mathcal H_\emptyset = \mathcal H_{CSCC}$.
\end{enumerate}

Note that, by property (ii), any $J_\Sigma \in \mathcal J_\Sigma$ can be uniquely written in the form $J^0_\Sigma \sqcup J_\emptyset$ where $M[J_\emptyset]$ has no non-dynamical CS-boundary  and where each connected component of  $M[J^0_\Sigma]$ has a non-trivial non-dynamical CS-boundary associated with a connected component of $\Sigma$.  In particular, $J_\emptyset\in \mathcal J_\emptyset$ and $J^0_\Sigma \in \mathcal J_\Sigma$.  Furthermore, for $J_i \in \mathcal J_\emptyset$, suppose that $|\psi_\emptyset\rangle := \sum_i c_i |J_i\rangle$ is a null state.  Then for $J^i_\Sigma = J^0_\Sigma \sqcup J_i$ and any $J'{}_\Sigma \in \mathcal J_\Sigma$, we
can consider $|\psi_\Sigma\rangle : = \sum_i c_i |J^i_\Sigma\rangle$ and compute
\begin{eqnarray}
\langle J'_\Sigma | \psi_\Sigma \rangle &=& \sum_i c_i \langle J'_\Sigma | J^i_\Sigma \rangle
\cr &=& \sum_i c_i \zeta[  J'_\Sigma{}^* \cup_\Sigma J^i_\Sigma ]
\cr &=& \sum_i c_i \zeta[ \left( J'_\Sigma{}^* \cup_\Sigma J^0_\Sigma \right) \sqcup J_i ]
\cr &=& \sum_i c_i \langle \left( J'_\Sigma{}^* \cup_\Sigma J^0_\Sigma \right)^* | J_i \rangle
\cr &=& \langle \left( J'_\Sigma{}^* \cup_\Sigma J^0_\Sigma \right)^* | \psi_\emptyset \rangle
 =0;
\end{eqnarray}
i.e., if $|\psi_\emptyset \rangle$ is a null state, then so is the state $|\psi_\Sigma\rangle$ defined by inserting the half-boundary $J_\Sigma^0$ in the boundary conditions that define every term $|\psi_\emptyset \rangle$.   As a result, for any
 $|\phi\rangle_{CSCC} \in \mathcal H_{CSCC}$ we can define an associated state $|J^0_\Sigma; \phi_{CSCC}\rangle \in \mathcal H_\Sigma.$ We may then denote the space of all such states with fixed $|\phi\rangle_{CSCC}$ by $\mathcal H_\Sigma^\phi$.  In particular, taking $\phi$ to range over the $\alpha$-states $|\alpha\rangle$ yields
\begin{equation}
\label{eq:HSigassumoveralpha}
\mathcal H_\Sigma = \oplus_\alpha \mathcal H_\Sigma^\alpha,
\end{equation}
where we have used completeness of the $\alpha$-state basis, the superselection rules $\widehat{Z[J]}|\alpha \rangle = Z_\alpha[J] |\alpha\rangle$, and orthogonality of distinct $\alpha$-states.  We can thus rewrite \eqref{eq:HQG1} in the form
\begin{equation}
\mathcal H_{QG} =
\oplus_{\alpha} \mathcal H_{QG}^\alpha \ \ \ {\rm with} \ \ \  \mathcal H_{QG}^\alpha =
\oplus_{\Sigma} \mathcal H_{\Sigma}^\alpha.
\end{equation}
Since the label $\alpha$ is superselected, we can then think of $\mathcal H_{QG}$ as an ensemble of Hilbert spaces labelled by $\alpha$.  In particular, the states  $|J^0_\Sigma; \alpha_{CSCC}\rangle$ provide a meaningful sense in which we can interpret every member of the ensemble as being defined by an $\alpha$-independent space of states (spanned by states of the form $|J^0_\Sigma\rangle$) on which we then define an $\alpha$-dependent inner product.  However, it is important to understand that this inner product will generally lead to an $\alpha$-dependent space of null states, so that the associated equivalence classes of states (and thus the dimension of the Hilbert space $\mathcal H_\Sigma^\alpha$) {\it do} generally  depend on $\alpha$.

As a final comment, note that the manner in which a given Hilbert space $\mathcal H_{\Sigma}^\alpha$ was defined above makes it more than an abstract Hilbert space.  For each $J \in \mathcal J$, our $\mathcal H_{\Sigma}^\alpha$ also knows that the operator $\widehat{Z[J]}$ acts as $Z_\alpha[J] {\mathds 1}$ on every state of $\mathcal H_{\Sigma}^\alpha$.  This is, of course, also true when $\Sigma = \Sigma_1\sqcup \Sigma_2$.  Since $\sigma[\mathcal J]$ satisfies property (1), it is then clear that the tensor product
$\mathcal H_{\Sigma_1}^\alpha \otimes \mathcal H_{\Sigma_2}^\alpha$ is always contained in $\mathcal H_{\Sigma_1\sqcup \Sigma_2}^\alpha$ though, as we will see in section \ref{subsec:noHfact}, the latter generally contains additional states.  On the other hand, the construction $\mathcal H_{\Sigma_1}^\alpha \otimes \mathcal H_{\Sigma_2}^{\alpha'}$ for $\alpha \neq \alpha'$ cannot arise as the action of the operator $\widehat{Z[J]}$ on such a space would be ill-defined.

\section{Model and results}
\label{sec:results}
The framework of section \ref{sec:framework} was used in \cite{Marolf:2020xie} to study simple topological models of quantum gravity in two bulk dimensions.  We now briefly remind the reader of both the models and the main results.  The models were parameterized by a number $k$ of end-of-the-world-brane (EOW brane) flavors, as well as coupling constants $S_0$ and $S_\partial$.  For simplicity, we will focus on the case $k=0$ (with no EOW branes) in the main text.  However, the case with EOW branes is briefly described in appendix \ref{subsec:withEOW}, along with a generalization that includes a third coupling constant $S_{EOW}$.

\subsection{Model without EOW branes}
\label{eq:noEOW}

The desired model is topological and contains no matter fields, so the allowed boundary conditions $J$ are entirely specified by their manifold structure $M[J]$ and (for this particular model) a choice of orientation; i.e., the orientation is the only source field, and the path integral does not depend on the choice of any boundary metric. Furthermore, since the bulk has dimension $2$, and since there are no EOW branes, the allowed $M[J]$ are closed one-dimensional manifolds. Any $J$ is thus just a collection of $n$ oriented circles.

The associated path integral is defined to be
\begin{equation}
\label{eq:EPImodel}
    \zeta[J] =\int_{\Phi \sim J} \mathcal{D} \Phi e^{-S[\Phi]} : = \sum_{{\mathcal M} \sim J} \mu(\mathcal M)e^{-S(\mathcal M)},
\end{equation}
where the action is defined in terms of the Euler character $\chi(\mathcal M)$ and the number of connected components $|\partial \mathcal M|$ of $\partial \mathcal M$ by choosing two parameters $S_0, S_\partial$  with $S_0 >0$ and writing
\begin{equation}
\label{eq:Smodel}
S(\mathcal M) = -S_0 \chi(\mathcal M) -S_\partial |\partial \mathcal M|.
\end{equation}
The measure factor in \eqref{eq:EPImodel} is taken to be
\begin{equation}
\label{eq:mumodel}
\mu(\mathcal M) = \frac{1}{\prod_g m_g!},
\end{equation}
where $m_g$ is the number of connected components of $\mathcal M$ with genus $g$ that have no asymptotic boundary.  As described in \cite{Marolf:2020xie}, this measure is the natural result of treating compactly-supported diffeomorphisms as gauge symmetries and thus dividing by the volume of the associated gauge group.

\subsubsection{Cauchy-surface compact cosmological States}

Let us use $J_\lcirclearrowright$ to denote the boundary condition defined by a single circle.
Since all allowed boundary conditions are disjoint unions of $J_\lcirclearrowright$, by \eqref{eq:Zcom} the corresponding operators $\widehat{Z[J]}$ on CSCC states are all powers of the single operator $\widehat{Z[J_\lcirclearrowright]}$.  In order to simplify the notation, we thus define $\hat Z :=  \widehat{Z[J_\lcirclearrowright]}$ and write any $\widehat{Z[J]}$  as $\hat Z^n$ for some $n$.  Explicit computations in \cite{Marolf:2020xie} then give the moments
\begin{equation}
\label{eq:Zmoments}
\frac{\langle \emptyset |\hat Z^n | \emptyset \rangle }{\langle \emptyset | \emptyset \rangle } = \frac{\zeta\left[\sqcup_{i=1}^n J_\lcirclearrowright \right]}{\zeta[\emptyset ] }  = \sum_{d=0}^\infty \left(e^{S_\partial - S_0}d\right)^n p_d
\end{equation}
with
\begin{equation}
\label{eq:lambda}
p_d = e^{-\lambda} \frac{\lambda^d}{d!} \ \ \ {\rm and} \ \ \ \lambda = \frac{e^{2S_0}}{1 - e^{-2S_0}}.
\end{equation}
Because the series that defines $\langle \emptyset |e^{iu\hat Z} | \emptyset \rangle$ converges, one can take the inverse Fourier transform of the result to show that (as one would expect from \eqref{eq:Zmoments}) the spectrum of $\hat Z$ is given by $e^{S_\partial - S_0}$ times the natural numbers $d \in {\mathds Z}^+ \cup \{0\}$, and that the no-boundary state $|\emptyset\rangle$ defines a probability measure on the associated eigenvalues given by the above Poisson distribution $p_d$.  We may thus denote the associated normalized eigenstates by $\Big| Z = e^{S_\partial - S_0} d\Big\rangle$ and write
\begin{equation}
\frac{\Big| \Big\langle  Z = e^{S_\partial - S_0} d | \emptyset \Big\rangle \Big|^2 }{\langle \emptyset| \emptyset \rangle } = p_d.
\end{equation}
The states $\Big| Z = e^{S_\partial - S_0} d\Big\rangle$ are the $\alpha$-states of this model, and for this theory we can replace the general label $\alpha$ by the more specific label $d\in {\mathbb Z} \cup \{0\}$, writing $|d\rangle := \Big| Z = e^{S_\partial - S_0} d\Big\rangle.$

For later use, we note that it is easy to explicitly construct the eigenstates $\Big| Z = e^{S_\partial - S_0} d\Big\rangle$ by simply choosing an entire function\footnote{We remind the reader that entire functions are those for which the series representation converges on the entire complex plane.} $f_d$ which vanishes at all arguments $e^{S_\partial - S_0} n$ for non-negative integers $n\neq d$ and for which $f_d(e^{S_\partial - S_0}d)=1$.  The operator $f_d(\hat Z)$ is then a projection onto the eigenstate $\Big| Z = e^{S_\partial - S_0} d\Big\rangle$ .  For example, we may choose
\begin{equation}
f_d = \frac{\sin (\pi e^{-{(S_\partial - S_0)}}\hat Z)}{\pi\left((e^{-{(S_\partial - S_0)}}\hat Z - d\right)}
\end{equation}
and write the normalized eigenstates in the form
\begin{equation}
\label{eq:alphastatesd}
\Big|   Z = e^{S_\partial - S_0} d \Big\rangle = \left(\frac{\lambda^d}{d!}\right)^{-1/2} \frac{\sin (\pi e^{-{(S_\partial - S_0)}}\hat Z)}{\pi\left((e^{-{(S_\partial - S_0)}}\hat Z - d\right)} |\emptyset \rangle .
\end{equation}
The reader should note that the boundary condition $\frac{\sin (\pi e^{-{(S_\partial - S_0)}}\hat Z)}{\pi\left((e^{-{(S_\partial - S_0)}}\hat Z - d\right)}$ used to define the $\alpha$-state \eqref{eq:alphastatesd} depends not only on the eigenvalue $Z_\alpha = e^{-{(S_\partial - S_0)}}d$ (or on the integer $d$), but also on the couplings $S_0, S_\partial$ as well.   Indeed, it was built by taking into account the full ($S_0$- and $S_\partial$-dependent) spectrum of $\hat Z$ and selecting the particular eigenvalue $Z_\alpha$. These states are all of the $\alpha$-states of the model without EOW branes.

The above results are not new.  However, we wish to emphasize that $p_d$ is positive for all values of $S_0, S_\partial.$  As a result, the inner product on the above CSCC states is {\it always} positive definite and $\mathcal H_{CSCC}$ is a Hilbert space, as desired.

\subsubsection{States with asymptotic boundaries}

Let us now consider the spaces of states $\mathcal H_\Sigma$ for non-trivial  $\Sigma$.  Since the allowed CS-boundaries $\Sigma$ must be given by slicing open a set of oriented circles, each $\Sigma$ is just some even number $2n$ of oriented points (or, equivalently, points that are assigned either a $+$ or $-$ sign), with $n$ of the points having positive orientation and $n$ of the points having negative orientation.  We may thus denote these spaces by $\mathcal H_{n,n}$.

As described in section \ref{subsec:Sigma}, states in $\mathcal H_{n,n}$ are defined by considering `half boundary conditions.'  We will use the description of this space associated with \eqref{eq:HSigassumoveralpha} where we describe states by choosing an $\alpha$-state in $\mathcal H_{CSCC}$
 and a $J^0_\Sigma$ for which every connected component has a non-trivial boundary (associated with a connected component of $\Sigma$).  We have seen that each $\alpha$-state is labelled by a non-negative integer $d\in {\mathds Z}^+ \cup \{0\}$.  Furthermore,   since $\Sigma$ is just a collection of $n$ positive points and $n$ negative points, and since each component of $J^0_\Sigma$ is an oriented half-circle (and is thus topologically equivalent to any directed curve), the allowed $J^0_{n,n} := J^0_\Sigma$ are defined by the possible ways to map the $n$ positive points onto the $n$ distinct negative points. If we denote the set of such $J^0_{n,n}$ by $\mathcal J^0_{n,n}$, then we see that $\mathcal J^0_{2n}$ has order $n!$.  Similarly,  formal linear combinations of the states $|J^0_{n,n}\rangle$ define a vector space $V_{n,n}$ of dimension $n!$. In particular,  the vector space structure of $V_{n,n}$ has no dependence on $d$.

However, we now wish to understand salient features of the $d$-dependent inner product on $V_{n,n}$.  One observation follows from Lemma 3 of \cite{Colafranceschi:2023urj}.   That Lemma considered the particular linear combination of the states $|J^0_{n,n} \rangle \in V_{n,n}$ that is completely anti-symmetric under permutations of either the $n$ positive boundary points or the $n$ negative boundary points.  It then showed that the norm of this state is a positive constant multiplied by the product
$Z_d(Z_d-1)\dots(Z_d-n+1)$ where $Z_d= \zeta_d(J_\lcirclearrowright)  = e^{S_\partial - S_0} d$.  We thus see that this state has negative norm whenever e.g. $n-1> Z_d > n-2.$

On the other hand, one can also show that the inner product on $V_{n,n}$ is positive definite whenever the eigenvalue $Z_d$ of $\hat Z$ is sufficiently large.  To do so, recall that for $I \in \mathcal J^0_{n,n}$ the states $|I; d_{CSCC}\rangle$ (built from the $\alpha$-state $|d\rangle := \Big| Z = e^{S_\partial - S_0} d\Big\rangle$) span a dense subspace of $\mathcal H^d_{n,n}$ (i.e., of $\mathcal H^\alpha_{n,n}$ where $Z_\alpha = de^{S_\partial-S_0}$).   Recall also that any such
$I \in \mathcal J^0_{n,n}$ is a collection of $n$ oriented line-segments which organizes the $2n$ points of $\Sigma$ into $n$ pairs.  It is then easy to compute the corresponding $d$-dependent norm of any such state, since each line segment in the bra-vector joins with the corresponding line segment in the ket-vector to make an oriented circle $J_\lcirclearrowright$.  Thus we find
\begin{equation}
\label{eq:Inorm}
\langle I; d_{CSCC} | I; d_{CSCC} \rangle = \zeta_d\left(\sqcup_{i=1}^n J_\lcirclearrowright\right) = e^{n(S_\partial - S_0)} d^n.
\end{equation}

The norms \eqref{eq:Inorm} are manifestly positive.  But we wish to study the norm of a general
state $|\psi; d_{CSCC}\rangle \in \mathcal H^d_{n,n}$ given by a linear combination of our basis states:
\begin{equation}
|\psi; d_{CSCC} \rangle = \sum_{I\in  \mathcal J^0_{n,n}} c_I |I; d_{CSCC}\rangle.
\end{equation}
Its norm is then
\begin{eqnarray}
\label{eq:psinorm}
\langle \psi; d_{CSCC}|\psi; d_{CSCC}\rangle &=& \sum_{I,J \in  \mathcal J^0_{n,n}} c^*_J c_I \langle J; d_{CSCC} |I; d_{CSCC}\rangle \cr
&=& \sum_I |c_I|^2 Z_d^n + \sum_{I\neq K \in  \mathcal J^0_{n,n}} c^*_K c_I \langle K; d_{CSCC} |I; d_{CSCC}\rangle,
\end{eqnarray}
where the first term in the second line comes from the diagonal inner products $\langle I; d_{CSCC} | I; d_{CSCC} \rangle$. The off-diagonal inner products $\langle K; d_{CSCC} | I; d_{CSCC} \rangle$  with $K\neq I$ are not zero, and they clearly play an important role in obtaining a negative norm for the totally anti-symmetric state for non-integer $Z_d$ and appropriate values of $n$.  However, since the line segments in the bra and ket no longer pair perfectly, off-diagonal inner products $\langle K; d_{CSCC} | I; d_{CSCC}\rangle$ with $K\neq I$ always involve strictly {\it less} than $n$ circles.  They thus represent functions of $Z_d$ that grow less rapidly than $Z_d^n$ at large $Z_d$.

 Since $c^*_K c_I \leq \sum_I |c_I^2|$, we also have
\begin{equation}
\sum_{I\neq K}|c^*_K c_I| \leq (n!)^2\sum_I |c_I^2|.
\end{equation}
As a result,  for any fixed coefficients $c_I$ and $Z_d> (n!)^2$, the (positive) first term on the 2nd line
 of \eqref{eq:psinorm} will be more important than the remaining term so that $\langle \psi; d_{CSCC}|\psi; d_{CSCC}\rangle > 0$ when any coefficient $c_I$ is non-zero.

 In other words, the inner product on $V_{n,n}$  must be positive definite for $Z_d> (n!)^2$.  Note that this is the case regardless of whether $Z_d$ is an integer.
In particular, for any choices of $S_0$ and $S_\partial$, and for any fixed $n$, the inner product on $V_{n,n}$ can fail to be positive definite only for finitely many values of $d$.  On the other hand, for fixed $d$, the inner product on $V_{n,n}$ will fail to be positive definite for infinitely many values of $n$ unless $Z_d \in {\mathds Z}^+ \cup \{0\}$.

\subsection{Failures of state space factorization}
\label{subsec:noHfact}

The physics of the topological model \eqref{eq:EPImodel} described above differs in several ways from that  associated with the standard AdS/CFT paradigm.  One such failure is that, even in a given $\alpha$ state $\Big|   Z = e^{S_\partial - S_0} d \Big\rangle$, for non-negative integers $n,m$ the space $\mathcal H^\alpha_{n+m,n+m}$ is not the tensor product  $\mathcal H^\alpha_{n,n}\otimes \mathcal H^\alpha_{m.m}$.  This failure of state-space factorization is also known as a failure of `Harlow factorization' due to the emphasis on this issue in \cite{Harlow:2015lma}.  For example, when $n=m=1$, the spaces $\mathcal H^\alpha_{n,n}$, $\mathcal H^\alpha_{m,m}$ each contain a single state, so that $\mathcal H^\alpha_{n,n}\otimes \mathcal H^\alpha_{m,m}$ is again one-dimensional. However,  as shown above, for large enough $Z_d$ there are no null states in $V_{n+m,n+m}=V_{2,2}$, so the Hilbert space $\mathcal H^\alpha_{n+m,n+m}=\mathcal H^\alpha_{2,2}$ has dimension $2! = 2$.

There is a bit more to say in the case where when $Z_d$ is a (nonnegative) integer.  In that case, we saw above that there were states in $V_{n,n}$ with norm proportional to $Z_d(Z_d-1)\dots(Z_d-n+1)$.  In particular, there are null states for $n> Z_d+1$.  In fact, for integer $Z_d$ we will argue in appendix \ref{subsec:withEOW} (see in particular footnote \ref{foot:nonIntY}) that the inner product is positive semi-definite on each $V_{n,n}$, so that $\mathcal H_{n,n}$ is a Hilbert space.  Noting that we can represent the identity operator on $\mathcal H_{n,n}$ by a collection of $2n$ line segments (where the operator acts by gluing the $m$th line segment to a given state at the $m$th boundary point) allows us to apply the argument of section 4.1 of \cite{Marolf:2020xie} to show\footnote{The argument in \cite{Marolf:2020xie} considered the identity on a one CS-boundary Hilbert space, but it applies directly to the identity on any $H_\Sigma$.} that the dimension of $\mathcal H_{n,n}$ is bounded by $Z_d^{2n}$.  For $n \gg Z_d^2$ this bound is far smaller than the dimension $n!$ of $V_{n,n}$, so most of the states in $V_{n,n}$ must be null.  We expect in this case that the dimension of $\mathcal H_{n,n}$ becomes precisely
$Z_d^{2n}$, and that for such large $n,m$ we will then have $\mathcal H_{n+m,n+m} = \mathcal H_{n,n}\otimes \mathcal H_{m,m}$.   Indeed, we expect that such large-$n$ restoration of state-space factorization will be a general feature of other models as well.  However, even for the current model, we leave detailed investigation of this issue for future work.

\section{Discussion}
\label{sec:disc}

The work above considered the $d=2$ topological gravity theories of \cite{Marolf:2020xie} for general values of the couplings $S_0, S_\partial$.  The main text studied the model without end-of-the-world branes (EOW branes), though appendix \ref{subsec:withEOW} considers models with EOW branes and in fact introduces an additional coupling $S_{EOW}$ not present in the analysis of \cite{Marolf:2020xie}.  While there were few truly new computations in the main text, we have endeavored to thoroughly describe the physics for all  values of $S_0, S_\partial, S_{EOW}$ whether or not the resulting Hilbert spaces were positive definite.    This is in contrast with the original treatment of \cite{Marolf:2020xie} which focused on values of the couplings that led to positive definite Hilbert spaces in all possible contexts.

The motivation for our more general analysis here was to use the general-coupling version of such systems as toy models that might suggest resolutions of the AdS/CFT factorization problem reviewed in the introduction.    Here we are interested in studying the bulk theory both at the mathematical level (where we simply wish to know what space of states and what inner product a given path integral might generate) and also at the physical level (where we ask to what extend such a theory might be physically viable).  At this physical level we will maintain the traditional perspective that viable theories should have a positive-definite Hilbert space.  However, we will not necessarily rule out a given set of couplings completely just because they lead to failures of positivity in certain sectors of the theory (which was the approach taken in \cite{Marolf:2020xie}).  Instead, we will adopt a more surgical approach in which we discuss the physical viability of a given set of couplings sector by sector, discarding only those sectors in which positivity fails.  The couplings will then be taken to describe a physically viable theory of any sectors that remain.

With this in mind, let us recall from the discussion in section \ref{sec:framework} that the framework of \cite{Marolf:2020xie} constructs an ensemble from the Cauchy-surface-compact cosmological (CSCC) sector of quantum gravity states, and in particular from the preferred basis of this space given by so-called $\alpha$-states.  An interesting observation is thus that, in the topological model discussed here without EOW branes, the inner product on this CSCC Hilbert space is positive definite for any values of $S_0, S_\partial$.  The associated ensemble is thus well-defined and elements of the ensemble are labelled by non-negative integers $d\in {\mathbb Z}^+ \cup \{0\}$, with each $d$ being assigned a non-negative probability $p_d$. As described in appendix \ref{subsec:withEOW}, the situation with EOW branes is more complicated and negative-norm zero-boundary states do in fact appear to arise when the quantity $Y_d = e^{S_{EOW} + S_\partial - S_0} d$ is not an integer.  But we will assume that this issue can be resolved by cutting out those parts of the theory in which positivity fails, and that this removal does not significantly affect interesting computations performed in appropriate no-boundary states $|\emptyset\rangle$.  In particular, in the model of appendix A, by a judicious choice of couplings and/or by discarding certain values of $d$ the theory can be restricted to cases with integer $Y_d$.

As described in the main text, the $\alpha$-states of the cosmological sector then define the Hilbert space structure of the sectors with non-trivial CS-boundaries.  However, the point we wish to emphasize is that, in the topological models studied here, the $\alpha$-states of the positive-definite CSCC just described generally still lead to {\it failures} of positivity in the presence of non-trivial CS-boundaries.  This is thus a context in which the above-mentioned surgical approach can be implemented, taking a large class of $\alpha$-states to be physically viable within the CSCC sector of the theory even when few of them are physically viable in sectors with non-trivial CS-boundaries.

The surgical approach has interesting implications for both the CSCC sector and the sectors with non-trivial CS-boundaries.  In particular, in the CSCC sector it provides a potential resolution of the cosmological decoherence problem in the introduction.  Recall that the problem was that,   in states of the theory that describe spacetimes with Cauchy surfaces that contain both compact and non-compact components, the part of the state describing closed cosmologies appears to be in constant danger of being permanently decohered by processes involving the non-compact components of the Cauchy surface; see again figure \ref{fig:timelabels}.    But this issue disappears for $\alpha$-parameters for which sectors of the Hilbert space with non-trivial CS-boundaries have been removed (say, due to failures of positivity)\footnote{One could equivalently say that there was never any such problem for the pure CSCC sector ${\cal H}_{CSCC}$, in which CS-boundaries are simply not allowed.}.

Nevertheless, as noted in footnote \ref{foot:dec} of the introduction, it remains to construct observables for the CSCC-sector that are sensitive to superpositions of $\alpha$-states.  At one level this is straightforward, as we need only consider operators on the CSCC Hilbert space that fail to be diagonal in the $\alpha$-basis.  Recall, however, that all observables constructed by inserting asymptotic boundaries (say, in the far past or far future) into the Euclidean path integral will necessarily act diagonally on $\alpha$-states due to the effect of wormholes.  Suppose then that we instead wish to act with the non-diagonal operator $|\alpha_1 \rangle \langle \alpha_2| $ defined by two distinct $\alpha$-states $|\alpha_1 \rangle$, $|\alpha_2 \rangle$, and let us in particular compute the matrix element  of the operator $|\alpha_1 \rangle \langle \alpha_2|$ between two states, $|\phi\rangle$ and $|\psi\rangle$:
\begin{equation}
\label{eq:nointworm}
\langle \phi  \Biggl( |\alpha_1 \rangle \langle \alpha_2 \Biggr) | \psi \rangle = \Biggl( \langle \phi   |\alpha_1 \rangle \Biggr) \Biggl( \langle \alpha_2  | \psi \rangle \Biggr).
\end{equation}
On the right-hand side, each of the factors in parenthesis can be computed using the path integral $\zeta$ that defines the theory, which in particular sums over all possible wormholes.  Thus the right-hand side is clearly a product of two path integrals.   The interesting point is then that if one wishes to instead describe \eqref{eq:nointworm} by a {\it single} path integral, one must declare it to be a path integral that (by definition!) does not include wormholes that connect the two factors in parentheses with each other\footnote{Instead allowing all such wormholes would make the path integral vanish as the boundaries that one inserts to obtain $|\alpha_1\rangle, |\alpha_2\rangle$ in fact create orthogonal projections onto these states that, with the inclusion of the above wormholes, become multiplied together.}.

Let us then turn to the implications for sectors of the theory with non-trivial (and perhaps asymptotically AdS) CS-boundaries.  In the presence of large numbers of such boundaries, for quantum gravity theories in sufficient numbers of dimensions it is plausible  that positivity of the Hilbert space might select a single member of the dual ensemble (which, in particular, might be a standard local CFT).  However, with regard to any properties of the theory that are uncorrelated with this positivity, this CFT would still behave as a typical member of the full ensemble.  In particular, such properties would be well-approximated by computations that average over the full ensemble.

We will also remark briefly on the connection to recent work in AdS/CFT suggesting that the CFT states dual to black holes have pseudorandom properties \cite{Cotler:2016fpe,Saad:2018bqo,Schlenker:2022dyo,Chang:2024zqi}.  In the topological model studied here, one can see from \eqref{eq:alphastatesd} that each $\alpha$-state (say, as defined by fixing $d$) depends smoothly on the couplings $S_0, S_\partial$.  It is thus natural to expect similar behavior in more general theories.  However, we also see that the Hilbert space defined by any fixed $d$ is positive definite only when $e^{-(S_\partial-S_0)}d$ is a non-negative integer.  This is a much more fragile condition.  It is thus natural to believe that, in more complicated models that include some degree of chaos, any values of $\alpha$ which give full positivity at given couplings will be pseudorandom.  In other words, we find it plausible that the pseudorandomness seen in CFT constructions could arise in the bulk from the selection of small numbers of $\alpha$-states (chosen to define positive definite Hilbert spaces) out of a large ensemble of more general elements.

In these senses, the `surgical' approach to $\alpha$-states advocated above --  and in particular the suggestion that the CSCC Hilbert space contains $\alpha$-states that fail to be physically viable in the presence of CS-boundaries -- would readily answer certain outstanding questions.  However, it also raises new issues that remain to be investigated.  One of these involves the fact that positivity is a rather strong property which can imply other constraints on the theory.  One might thus wonder which other properties of the theory are truly uncorrelated with positivity and can thus be reproduced by ensemble averages.

Another issue arises from the observation that, at large $d$ in our topological model without EOW branes, the failures of positivity typically required correspondingly large numbers of CS-boundaries.  In such cases, should we still consider sectors with small-but-nonzero numbers of CS-boundaries to be physically viable if their Hilbert spaces are positive definite?

On the other hand,  it is typically argued that contexts with multiple CS-boundaries (i.e., ``multiple separate universes'') should be a good model for computations involving multiple experiments in the same universe (i.e., in a context with a single CS-boundary) when the experiments are well-separated in space and time.  One might thus ask if the failures of positivity seen in the topological model at large numbers of CS-boundaries should be taken to suggest that models with local degrees of freedom will in fact be subject failures of positivity in contexts with a single CS-boundary (or perhaps even in cosmological contexts with no CS-boundaries) that involve large numbers of experiments that are  well-separated in space and time.  If so, it will clearly be important to understand the resulting implications for the surgical approach advocated above\footnote{It is interesting to note that negative-norm states contribute negatively to any partition function $Tr(e^{-\beta H})$, so that a positive density of negative-norm states would be indistinguishable from a negative density of positive-norm states.  One might wonder if this might somehow be connected to results of .  However, a complication is that the negative-norm states described in the main text are closely associated with a failure of Hilbert spaces to factorize over disconnected boundaries, and thus also with a failure of path-integral-partition functions to in fact give $Tr(e^{-\beta H})$. I.e., in a 2d theory the path integral over a disk of boundary length $\beta$ need not agree with $Tr(e^{-\beta H})$ where the trace is taken over the 1-boundary Hilbert space.}.

In particular, the discussion above has largely assumed that, when positivity fails in the sector ${\cal H}_\Sigma$ with a given CS-boundary $\Sigma$,  we would discard the sector  ${\cal H}_\Sigma$ as a whole (rather than attempting to `repair' it by choosing a sense in which to discard `only the states with negative norm').  But a more subtle approach may be needed in which individual states are in fact removed from a given sector.  As illustrated by the familiar example of 1+1 Minkowski space (in which no spacelike plane is preferred), there is generally no preferred way to truncate an inner product space of indefinite signature to one that is positive definite without introducing additional structure.  However, for the CSCC sector it is plausible that the $\alpha$-states themselves could provide a useful such structure in the sense that one would either keep or remove states with definite values of $\alpha$.  However, this remains to be investigated in detail\footnote{Such a framework would be mathematically similar to that associated with the contour-rotation prescription of \cite{Marolf:2022jra} (studied in more detail in \cite{Liu:2023jvm}), though where the negative-norm eigenvectors are now projected out rather than being Wick-rotated.}.

Finally, it may be interesting to revisit the physical questions raised by black hole evaporation in CSC-cosmologies under the above scenario.  If we carefully collect the Hawking radiation from very small but well-separated identically-prepared evaporating black holes, it is natural to expect that expectation values of swap operators in such joint-radiation states will continue agree with replica wormhole calculations.  In that case, the evaporation of such black holes would follow a Page curve, and would appear unitary as described by observers in cosmological spacetimes.  However, since we now allow $\alpha$-sectors that might be forbidden when considering sectors with CS-boundaries, it is interesting to ask whether a complete measurement of the state of Hawking radiation produced by such black holes in the cosmological context would yield results that differ in detail from those found for black holes with corresponding local environments in sectors with non-trivial CS-boundaries.  Furthermore, since the parameters influencing the final state of black hole evaporation in the cosmological sector of such a scenario are no longer in-principle superselected, they will in some cases be subject to interesting interference effects.  We look forward to exploring such issues further in future work.

\section*{Acknowledgements}

DM thanks Xi Dong, Steve Giddings, Gary Horowitz, Clifford Johnson, Henry Maxfield, Alex Maloney,  Misha Usatyuk, and Ying Zhao for useful discussions.  He also thanks Hirosi Ooguri, Eva Silverstein, Edward Witten, and the participants of the 2024 ``What is string theory?'' KITP program for comments on the presentation of various ideas in this work.
This research was supported in part by grant NSF PHY-2309135 to the Kavli Institute for Theoretical Physics (KITP). It was also was supported by NSF grant PHY-2107939, and by funds from the University of California.

\appendix

\section{Adding EOW branes}
\label{subsec:withEOW}

The topological model of \cite{Marolf:2020xie} can be extended by adding $k$ flavors of EOW branes, with branes of a given flavor treated as indistinguishable bosons. Having discussed the $k=0$ case in detail in the main text, we now rather briefly describe the case with EOW branes.  Our goal here is merely to extract the relevant results from \cite{Marolf:2020xie} (including a slight generalization of the couplings considered there) to justify the brief comments at the end of section \ref{sec:disc}.  We will include enough details so that, by comparing the comments below with the EOW-brane computations in \cite{Marolf:2020xie}, the intrepid reader will be able to reproduce the stated results.

With a non-trivial set of EOW branes,   the allowed boundary conditions $\mathcal J$ for the path integral will consist of some number of (unlabeled) oriented circles as in the original model, together with a set of (oriented) line segments with each endpoint labelled by some flavor of EOW branes.  We will again use $\hat Z$ to denote the operator on $\mathcal H_{CSCC}$ defined by adding circles to the boundary condition, and we will use $\widehat{(\psi_I,\psi_K)}$ to denote the operator on $\mathcal H_{CSCC}$ that adds a line segment running from EOW brane flavor $I$ to flavor $K$.  The relevant path integral is a simple extension of \eqref{eq:EPImodel}, which now allows the bulk manifold to have `EOW-brane boundaries.'  We refer the reader to \cite{Marolf:2020xie} for details.

However, it will be useful to introduce an additional coupling $S_{EOW}$ that was not studied in \cite{Marolf:2020xie}.  Recall that the coupling $S_\partial$ weights all circular boundaries (by a factor of $e^{S_\partial}$.  We will now weight any circular boundary that contains an EOW by an additional factor of $e^{S_{EOW}}$.

The addition of EOW branes allows us to define a non-trivial space of `half boundary conditions' $\mathcal J_{\Sigma}$ for any $\Sigma$ that consists of $n$ positively-oriented points and $m$ negatively-oriented points for any non-negative integers $n,m$.  We thus call the associated space of half boundary conditions $\mathcal J_{n,m}$.  The allowed boundary conditions attach oriented line segments to the boundary points according to the following rules:  Since the segments are oriented, one end of each segment is positive and one end is negative, with the  direction of the line segment running from the negative end to the positive end.
We begin by considering $n$ such segments and attaching their positive ends to the $n$ positively-oriented boundary points.  We then attach the corresponding negative ends one by one
{\it either} to an unconnected negative boundary point or to a label given by some EOW flavor $I_i$ for the $i$th such segment. In general, this will leave some number $m' < m$ of negative boundary points remaining unconnected.  We then attach them to the negative ends of $m'$ additional oriented line segments whose positive ends we again label with EOW flavors (say $K_j$ for the $j$th such segment).

 \begin{figure}[h!]
        \centering
\includegraphics[width=0.5\linewidth]{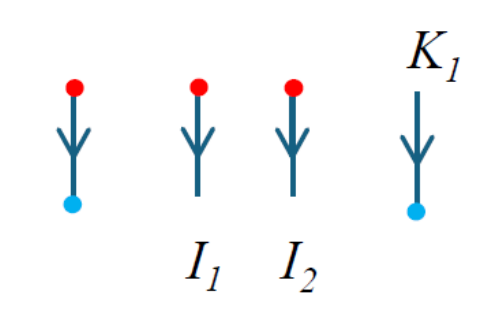}\caption{A simple example of a half boundary condition in $\mathcal J_{3,2}$, meaning that $\Sigma$ consists of 3 positively-oriented points (red) and 2 negatively-oriented points (blue).  In this case, the first positive point is connected to the first negative point, while the 2nd and 3rd positive points source EOW branes of flavors $I_1$ and $I_2$ and the 2nd negative point provides a source for an EOW anti-brane of flavor $K_1$.  The EOW branes and anti-branes are distinguished in the figure by the orientations of the associated line segments.}
\label{fig:EOWhalfBCs}
\end{figure}

It turns out that the introduction of EOW branes modifies the expectation values of $\hat Z^n$ in the no-boundary state $|\emptyset\rangle$ only by a change in expression \eqref{eq:lambda} for the parameter $\lambda$.  This changes the probabilities $p_d$ for any given eigenvalue $Z_d$ of $\hat Z$, but it leaves unchanged the {\it spectrum} of $\hat Z$ and the construction of the associated $\alpha$ states \eqref{eq:alphastatesd}. In particular, the new parameter $S_{EOW}$ has no effect on this spectrum.  It also turns out that the only change in $p_d$ is to alter the definition of the parameter $\lambda$ that controls the Poisson distribution to
\begin{equation}
\label{eq:newlambda}
\lambda = \frac{e^{2S_0}}{1 - e^{2S_0}}\left(1 + e^{S_{EOW}}\left[e^{ke^{S_\partial-S_0}}-1\right]  \right).
\end{equation}
Since \eqref{eq:newlambda} is again positive, the $p_d$ remain positive and they are still defined by non-negative integers $d$.   One can then consider individually each superselection sector defined by some $\hat Z$-eigenvalue $Z_d = e^{S_\partial - S_0} d$, compute correlators of the $\widehat{(\psi_I,\psi_J)}$ in this sector, and deduce properties of the corresponding distributions to find the corresponding spectrum of $\widehat{(\psi_I,\psi_J)}$.

While the coupling $S_{EOW}$ was not discussed in \cite{Marolf:2020xie}, we can nevertheless read off its effects directly from the analysis in that work. The upshot is that the
joint spectrum of the $\widehat{(\psi_I,\psi_J)}$ is now controlled by the quantity $Y_d:= e^{S_{EOW}}Z_d =
e^{S_{EOW} + S_\partial - S_0} d$ in the fashion we describe below\footnote{\label{foot:SEOW} Readers wishing to derive the results stated below should begin by noting that the role of $S^{EOW}$ is to multiply any boundary having at least one EOW brane by $e^{S^{EOW}}$.  As a result,  on the right-hand-side of equation (3.35) from \cite{Marolf:2020xie},  every term  in the argument of the exponential function that contains at least one factor of the matrix $t$ must now be multiplied by $e^{S_{EOW}}$. This observation allows one to see that the left-hand-side of (3.35) from \cite{Marolf:2020xie} becomes $e^{ue^{S_\partial}}/[\det(I-t)^{e^{(S_\partial +S_{EOW})}}]$, whence (3.37) from \cite{Marolf:2020xie} becomes $[\det(I-t)^{- Z_d e^{S_{EOW}}}]$.  Thus it is now the quantity $Y_d:= e^{S_{EOW}}Z_d =
e^{S_{EOW} + S_\partial - S_0} d$  that controls the spectrum of the operators $\widehat{(\psi_I,\psi_J)}$.}.

We will focus on the case where $Y_d$ is a non-negative integer, as this is the case where the inner product on the CSCC Hilbert space $\mathcal H_{CSCC}$ is positive definite\footnote{\label{foot:nonIntY} An exploration of the case where $Y_d$ fails to be a non-negative integer would largely follow the discussion in section 3.7 of \cite{Marolf:2020xie}.   For $k-1 < Y_d < k$ the inner product on $\mathcal H_{CSCC}$ remains positive definite and the joint spectrum of the $\widehat{(\psi_I,\psi_K)}$ is identical to that in the case $Y_d=k$.  However, for non-integer $Y_d < k-1$ the inner product on $\mathcal H_{CSCC}$ fails to be positive definite.}.  Clearly, however, this condition does {\it not} require the $Z_d$ to take integer values.  Nevertheless,   in this case
the joint spectrum of the $\widehat{(\psi_I,\psi_K)}$ consists of all positive semi-definite matrices $M_{IK}$ of rank $r \le \min(Y_d,k)$, with the case $r<\min(Y_d,k)$ having measure zero. A given $\alpha$-sector is defined by choosing some matrix $M_{IK}$ from this ensemble and taking the corresponding eigenvalue of $\widehat{(\psi_I,\psi_K)}$ to be this $M_{IK}$.  As a result, this matrix defines the (positive semi-definite) inner product $\langle I;+|K;+\rangle$ on the space $V_{1,0}$ of  states with one positive boundary, where $|K;+\rangle$ is the state in which the single positive point is attached to a line segment whose negative end is labelled with EOW brane flavor $K$.
Thus $V_{1,0}$ defines a Hilbert space $\mathcal H^\alpha_{1,0}$.  With probability one, the dimension of $\mathcal H^\alpha_{1,0}$ is $\min(Y_d,k)$.  Similarly, with a single negative boundary we find the Hilbert space $\mathcal H^\alpha_{0,1}$ defined by taking the inner product of the states $\langle I;- | K;- \rangle$ to be  $M_{IJ}^*$.

We may then consider the space $V_{1,1}$  defined by having one boundary of each sign.  Some of the states in $V_{1,1}$ contain a single positive-signed EOW brane and a single negative-signed EOW brane.  Such states span the space $V_{1,0} \otimes V_{0,1}$.  The inner product on this space takes the form $M\otimes M^*$, where $M^*$ is the complex conjugate of the matrix $M$.  (Since $M$ is self-adjoint, this $M^*$ is also the transpose of $M$.)   Such states thus define the Hilbert space $\mathcal H^\alpha_{1,0} \otimes \mathcal H^\alpha_{0,1}$.

While $V_{1,1}$ is not precisely $V_{1,0} \otimes V_{0,1}$, the only additional state in $V_{1,1}$ is the (oriented) semi-circle $|\cyl \rangle$. To better understand the role played by $|\cyl\rangle$, we may follow the argument of section 3.6 of \cite{Marolf:2020xie}. We thus begin by considering an orthonormal basis  $\{|i;+\rangle\}$ for $\mathcal H^\alpha_{1,0}$.  Note that we can express each
$|i;+\rangle$  in terms of the (not generally orthonormal) brane-flavor states $|K;+\rangle$ defined above by writing
\begin{equation}
|i;+\rangle = \sum_K a_{iK}|K;+\rangle.
\end{equation}
We may then introduce a corresponding set of states in $\mathcal H^\alpha_{0,1}$ defined by
\begin{equation}
|i;-\rangle = \sum_K a_{iK}^*|K;-\rangle,
\end{equation}
whence one can check from the above inner products that $\{|i;-\rangle\}$ is an orthonormal basis for $\mathcal H^\alpha_{0,1}$.

We can now also consider the state $|\nu \rangle : =|\cyl \rangle - \sum_i |i;+\rangle \otimes |i;-\rangle \in \mathcal H^\alpha_{1,1}$. Since the inner product of $|\cyl \rangle$ with
$|K;+\rangle \otimes |I;-\rangle$ is given by the eigenvalue of the operator $\widehat{(\psi_I, \psi_K)}$ in our $\alpha$-state, this inner product is $M_{IJ}$.  And since the $|i;+\rangle$ are normalized in
$\mathcal H^\alpha_{1,0}$, the inner product of $|\cyl \rangle$ with
$|i;+\rangle \otimes |i;-\rangle$ is just $1$.  A short computation then shows that
 the state $|\nu \rangle$ is orthogonal to the space $\mathcal H^\alpha_{1,0} \otimes \mathcal H^\alpha_{0,1}$.

Furthermore, for $k\ge  Y_d$ the state $|\nu \rangle$ has norm
\begin{equation}
\langle \nu |\nu \rangle = Z_d - 2 \min(Y_d,k) + \min(Y_d,k) =Z_d-\min(Y_d,k).
\end{equation}
As a result, for $Z_d = \min(Y_d,k)$ the additional state $|\nu\rangle$ is a null state and
we find $\mathcal H^\alpha_{1,1} = \mathcal H^\alpha_{1,0} \otimes \mathcal H^\alpha_{0,1}.$   A similar argument then holds for any number of boundaries so that in fact $\mathcal H^\alpha_{n,m}$ is the tensor product of $n$ copies of $\mathcal H^\alpha_{1,0}$ and $m$ copies of $\mathcal H^\alpha_{0,1}.$ This also supplies the proof of positivity mentioned in section \ref{subsec:noHfact} above\footnote{Section \ref{subsec:noHfact} in fact stated that each Hilbert space $\mathcal H^\alpha_{n,n}$ was positive definite for the case $k=0$.  But that Hilbert space is just the subspace of the above  $\mathcal H^\alpha_{n,n}$ defined by restricting $V_{n,n}$ to linear combinations of the half-boundary conditions that contain no EOW brane flavor-labels.  The desired claim then follows from the observation that, given any Hilbert space with positive-definite inner product, the inner product induced on any subspace is also positive-definite.}.

On the other hand, for $Z_d - \min(Y_d,k)>0$ we find that $|\nu\rangle$ is a new positive norm state.  Since it is the only state in $\mathcal H_{1,1}$ that is orthogonal to $\mathcal H_{1,0} \otimes \mathcal H_{0,1}$, the inner product on $\mathcal H_{1,1}$ is positive definite as one might desire.  But for $Z_d < \min(Y_d,k)$ the norm of $|\nu\rangle$ is negative and $\mathcal H_{1,1}$  is not in fact a Hilbert space.

\addcontentsline{toc}{section}{References}
\bibliographystyle{JHEP}
\bibliography{references}

\end{document}